\input{style/arxiv-general.cfg}
\documentclass[aoas,MSNbibl,nameyear,seceqn,dvips]{arximspdf}
\makeatletter
   \@ifpackageloaded{graphicx}{}{\usepackage{graphicx}}
\makeatother
\usepackage{algorithmic}
\usepackage{algorithm}

\doi{10.1214/14-AOAS794}% Updated by VTEXPTS2LaTeX.exe, 26.11.2014
\volume{9}
\issue{1}
\pubyear{2015}
\firstpage{324}
\lastpage{352}
\docsubty{FLA}

\makeatletter
\newcommand{\TT}{\mathrm{TT}}
\newcommand{\TP}{\mathrm{TP}}
\newcommand{\PP}{\mathrm{PP}}
\newcommand{\PT}{\mathrm{PT}}
\newcommand{\rrvert}{\vert}
\newcommand{\rrVert}{\Vert}
\newcommand{\llvert}{\vert}
\newcommand{\llVert}{\Vert}
\renewcommand{\mid}{|}
\def\tr{\operatorname{tr}}
\def\argmax{\operatorname{argmax}}
\def\MSE{\mathrm{MSE}}
\def\cov{\operatorname{Cov}}
\makeatother

\begin{document}
\begin{frontmatter}

\title{Statistical paleoclimate reconstructions via Markov random fields}
\runtitle{Paleoclimate reconstruction via MRF}

\begin{aug}
% Corresponding author: Dominique Guillot - dguillot@stanford.edu% Updated by VTEXPTS2LaTeX.exe, 26.11.2014 14:42
%Updated by VTEXPTS2LaTeX.exe, 26.11.2014 11:41
\author[A]{\fnms{Dominique}~\snm{Guillot}\corref{}\thanksref{M1,T1,T2}\ead[label=e1]{dguillot@stanford.edu}},
\author[A]{\fnms{Bala}~\snm{Rajaratnam}\thanksref{M1,T2}\ead
[label=e2]{brajarat@stanford.edu}}
\and
\author[B]{\fnms{Julien}~\snm{Emile-Geay}\thanksref{M2,T3}\ead[label=e3]{julieneg@usc.edu}}
\runauthor{D. Guillot, B. Rajaratnam and J. Emile-Geay}
\affiliation{Stanford University\thanksmark{M1} and University of
Southern California\thanksmark{M2}}
\address[A]{D. Guillot\\
B. Rajaratnam\\
Department of Statistics\\
Stanford University\\
Sequoia Hall\\
390 Serra Mall\\
Stanford, California 94305-4065\\
USA\\
\printead{e1}\\
\phantom{E-mail: }\printead*{e2}}
\address[B]{J. Emile-Geay\\
Department of Earth Sciences\\
\quad and Center for Applied Mathematical Sciences\\
University of Southern California\\
3651 Trousdale Parkway, ZHS 275\\
Los Angeles, California 90089-0740\\
USA\\
\printead{e3}}
\end{aug}
\thankstext{T1}{Supported in part by a NSERC
postdoctoral fellowship, and by funding from the Universtity of Southern California and
Stanford University.}
\thankstext{T2}{Supported in part by NSF
Grants DMS-0906392, DMS-CMG 1025465, AGS-1003823, AGS-1003818, DMS-1106642,
DMS-CAREER-1352656 and Grants DARPA-YFAN66001-111-4131, AFOSR
FA9550-13-1-0043, UPS fund and SMC-DBNKY.}
\thankstext{T3}{Supported in part by the National Science Foundation under Grant AGS-1003818.}

% HISTORY:
%
\received{\smonth{3} \syear{2013}}% Updated by VTEXPTS2LaTeX.exe,
%26.11.2014 11:41
%
\revised{\smonth{10} \syear{2014}}% Updated by VTEXPTS2LaTeX.exe,
%26.11.2014 11:41

% ABSTRACT
%
\begin{abstract}
Understanding centennial scale climate variability requires data sets
that are accurate,
long, continuous and of broad spatial coverage. Since instrumental
measurements are generally only available after 1850, temperature
fields must be reconstructed using paleoclimate archives, known as
proxies. Various climate field reconstructions (CFR) methods have been
proposed to relate past temperature to such proxy networks. In this
work, we propose a new CFR method, called GraphEM, based on Gaussian
Markov random fields embedded within an EM algorithm. Gaussian Markov
random fields provide a natural and flexible framework for modeling
high-dimensional spatial fields. At the same time, they provide the
parameter reduction necessary for obtaining precise and
well-conditioned estimates of the covariance structure, even in the
sample-starved setting common in paleoclimate applications. In this
paper, we propose and compare the performance of different methods to
estimate the graphical structure of climate fields, and demonstrate how
the GraphEM algorithm can be used to reconstruct past climate
variations. The performance of GraphEM is compared to the widely used
CFR method RegEM with regularization via truncated total least squares,
using synthetic data. Our results show that GraphEM can yield
significant improvements, with uniform gains over space, and far better
risk properties. We demonstrate that the spatial structure of
temperature fields can be well estimated by graphs where each neighbor
is only connected to a few geographically close neighbors, and that the
increase in performance is directly related to recovering the
underlying sparsity in the covariance of the spatial field. Our work
demonstrates how significant improvements can be made in climate
reconstruction methods by better modeling the covariance structure of
the climate field.\vspace*{25pt}
\end{abstract}

% KEYWORDS
\end{frontmatter}

\setcounter{footnote}{3}

%s1 #&#
\section{Introduction and preliminaries}
%s1.1 #&#
\subsection{Introduction}
\begin{figure}%

\includegraphics{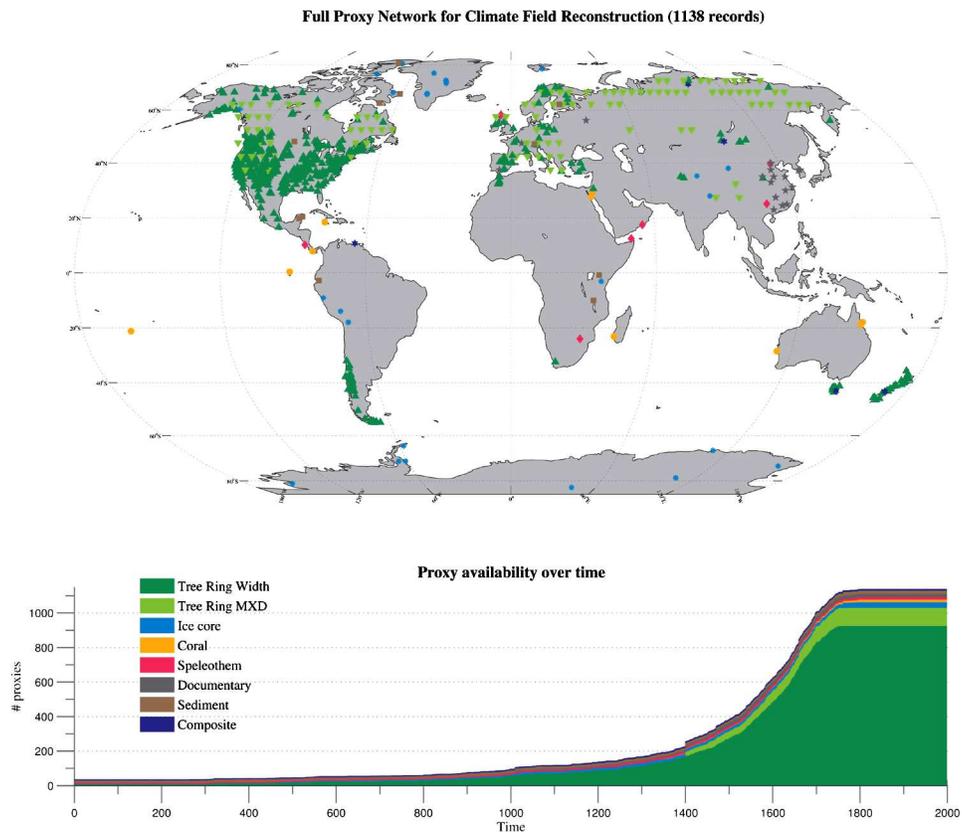}

\caption{Location, type and temporal availability of proxies in the
\citet{Mann08a} database.}
\label{figproxyavail}
\end{figure}

Fundamental to an informed quantification of recent climate change is
an accurate depiction of past climate variability [\citet
{AR5chap5}]. Since widespread instrumental observations of surface
temperatures are only available after the mid-nineteenth century,
climate scientists rely on proxy data (e.g., tree rings, ice cores,
sediment cores, corals) to infer past temperatures via statistical
modeling [\citet{JonesHolocene09,NRC06}]---a task known as
``paleoclimate reconstruction'' in the climate literature. Given an
instrumental temperature data set [see, e.g., \citet{Brohan06}]
and a global network of climate proxies [e.g.,~\citet{Mann08a},
Figure~\ref{figproxyavail}], the temperature back in time can be
estimated as a function of proxies.

Various CFR methods have been proposed to infer past climate [see
\citet{TingleyQSR2012}]. Here we adopt an approach based on
multivariate linear regression as in the regularized EM algorithm
[\citet{Schneider01}]. In that setting, the CFR problem is
formalized as a missing data problem, which we now describe.

\begin{figure}

\includegraphics{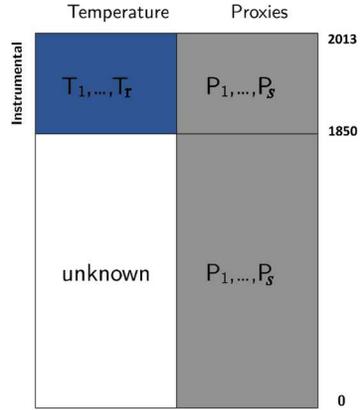}

\caption{Temperature/proxy matrix with missing values.}
\label{figdatamatrix}
\end{figure}

Consider a spatial grid and let $p$ denote the number of temperature
and proxy points. Let $n = n_a + n_m$ denote the sum of the number of
years of available instrumental data, $n_a$, and missing data, $n_m$.
In practice, $p \approx3000$, $n \approx2000$ and $n_a \approx150$
(instrumental period). We model the temperature and proxy points as a
multivariate random vector $(X_1, \ldots, X_p) \sim N_p(\mu, \Sigma)$
with missing values, where $\mu= (\mu_1, \ldots, \mu_p)$ is the mean
vector and $\Sigma= (\sigma_{ij})_{p \times p}$ is the covariance
matrix of the model. We denote by $X$ the (incomplete) $n \times p$
data matrix where each row represents a year of observations containing
$r$ instrumental temperature observations and $s$ proxy measurements.
Hence, the rows represent time order and the columns represent
different spatial locations of both instrumental temperature and proxy
data (see Figure~\ref{figdatamatrix}).
%f1 #&#
%

Figure~\ref{figproxyavail} shows that the availability of the proxy
data from the network of \citet{Mann08a} decreases rapidly in
time, and missing values constitute as much as $80\%$ of the entries in
the matrix.
%f2 #&#
%
Reconstructing the pre-instrumental temperature field may be cast as a
missing data problem, for which several strategies exist [\citet
{LittleRubin2002}]. However, the high dimensionality of the problem
(``large $p$, small $n$'') makes it challenging to apply standard
methods. For instance, it is well known that the sample covariance
matrix is a poor estimator of $\Sigma$ in that setting [\citet
{lin85,Paul07,stein77}]. In this paper, we explore the use of Gaussian
Markov random fields (a.k.a. Gaussian graphical models) for estimating
$\Sigma$. This approach provides flexibility in terms of modeling the
inherent spatial heterogeneities of the field, but at the same time
reduces the number of parameters that need to be estimated, thereby
leading to improved reconstructions of past temperature. We start by
recounting existing reconstruction strategies before introducing our
new approach.

%s1.2 #&#
\subsection{The EM algorithm}
A popular method for the imputation of missing values is the EM
algorithm [\citet{DempsterEM,LittleRubin2002}]. In the
multivariate normal setting, given an estimate of $\mu$ and $\Sigma$,
the EM algorithm reduces to regressing the missing values on the
available ones, and thereafter updating the estimates of $\mu$ and
$\Sigma$. This procedure is iterated until convergence. More
precisely, let $x$ denote the $k$th row of $X$, and let $x_a$ and $x_m$
denote the parts of $x$ where data are available and missing,
respectively. Let\vspace*{1pt} $\mu^{(0)}$ and $\Sigma^{(0)}$ be initial estimates
of $\mu$ and $\Sigma$. For example, $\mu^{(0)}$ and $\Sigma^{(0)}$
could be the sample mean and sample covariance of the data set
completed by replacing every missing value by the mean of the available
values in the corresponding columns of $X$ [\citet{Schneider01}].
The EM algorithm iteratively constructs a sequence $\mu^{(l)}$ and
$\Sigma^{(l)}$ of estimates of $\mu$ and $\Sigma$. For every $l \geq
0$, the E-step consists of a linear regression
%e1.1 #&#
%
\begin{equation}
\label{eqnregression} \bigl(x_m-\mu^{(l)}_m
\bigr)^\top= B^{(l)} \bigl(x_a-
\mu^{(l)}_a\bigr)^\top,
\end{equation}
where
%e1.2 #&#
%
\begin{eqnarray}
\label{eqnreg} B^{(l)} &=& \Sigma^{(l)}_{ma} \bigl(
\Sigma^{(l)}_{aa}\bigr)^{-1} , \qquad\Sigma
^{(l)} = \pmatrix{ \Sigma^{(l)}_{aa} &
\Sigma^{(l)}_{am}
\vspace*{3pt}\cr
\Sigma^{(l)}_{ma} &
\Sigma^{(l)}_{mm}}\quad\mbox{and}
\nonumber\\[-8pt]\\[-8pt]\nonumber
\mu^{(l)} &=& \bigl(
\mu^{(l)}_a, \mu^{(l)}_m\bigr),
\end{eqnarray}
are the regression coefficients and the decompositions of $\Sigma
^{(l)}$ and $\mu^{(l)}$ associated with the decomposition of $x$ among
its available and missing parts. Denote by $X^{(l+1)}$ the completed
estimate of $X$, obtained after the regression (\ref{eqnregression})
has been performed in order to impute the missing values in each row of
$X$. In the M-step of the algorithm, the estimates of $\mu$ and
$\Sigma$ are updated by
%e1.3 #&#
%
\begin{eqnarray}\label{eqnEMupdate}
\mu^{(l+1)}_i &=& \frac{1}{n} \sum
_{k=1}^n X^{(l+1)}_{ki},
\nonumber\\[-8pt]\\[-8pt]\nonumber
\Sigma^{(l+1)}_{ij} &=& \frac{1}{n} \sum
_{k=1}^n \bigl[\bigl(X^{(l+1)}_{ki}-
\mu^{(l+1)}_i\bigr) \bigl(X^{(l+1)}_{kj}-
\mu^{(l+1)}_j\bigr) \bigr] + C_{ij}^{(l+1)},\nonumber
\end{eqnarray}
where $C_{ij}^{(l+1)}$ is the covariance of the residuals. Using the
same block decomposition as in (\ref{eqnreg}), we have
%e1.4 #&#
%
\begin{equation}
\label{eqncovresidual} C^{(l+1)} = \pmatrix{ 0 & 0
\vspace*{3pt}\cr
0 & \Sigma^{(l)}_{mm}
- \Sigma^{(l)}_{ma} \bigl(\Sigma^{(l)}_{aa}
\bigr)^{-1}\Sigma^{(l)}_{am}}.
\end{equation}

The reader is referred to \citet{LittleRubin2002} and \citet
{McLachlanEM} for more details about the EM algorithm.

%s1.3 #&#
\subsection{The regularized EM algorithm}\label{secregill}
Obtaining a precise estimate of $\Sigma$ is a crucial step of the EM
algorithm. In the sample-starved setting common to many paleoclimate
problems, the sample covariance matrix is generally not invertible and
can be a very poor estimator of $\Sigma$. This is a serious problem
since parts of $\Sigma$ need to be inverted to compute the regression
coefficients $B$. Different $\ell_2$-type methods to regularize the
problem have been proposed in the literature. Among them are ridge
regression [a.k.a. Tikhonov regularization, \citet
{HankeHansen93,ESL08}, \citeauthor{Hoerl70a} (\citeyear{Hoerl70a,Hoerl70b}), \citet{Tikhonov77}] and truncated total
least squares [TTLS, \citet
{Fierro97,GolubVanLoan80,vanHuffelVandewalle91}] regression. These
methods can be used to replace the regression matrix $B^{(l)}$ in
equation (\ref{eqnregression}) by a regularized estimate, and have
been implemented within the EM algorithm. The resulting algorithm is
known as RegEM [\citet{Schneider01}] and has been widely used in
paleoclimate studies [\citeauthor{JEG10a} (\citeyear{JEG10a,JEG10b}),
\citeauthor{MRWA05} (\citeyear{MRWA05,MRWA07,Mann08a,Mann09}),
\citeauthor{Riedwyl08} (\citeyear{Riedwyl08,Rutherfordetal05})].
For example, in RegEM-TTLS, the linear regressions in the EM algorithm
are replaced by truncated total least squares (TTLS) regressions. The
TTLS solution of a linear system $Ax = b$ is obtained by expressing the
total least squares solution of the linear system as a function of the
SVD of the matrix $A$, and then truncating all but a given number of
eigenvalues. The number of retained eigenvalues corresponds to the \textit{truncation parameter} of RegEM-TTLS [see \citet{Fierro97} for
more details].

To date, all direct regression methods have resulted in reconstructions
that underestimate the amplitude of past climate variations to some
extent [e.g., \citeauthor{Smerdonetal2010} (\citeyear{Smerdonetal2010,SmerdonGRL11}),
\citet{vonStorch04}].
This ``regression dilution'' [\citet{FrostThompson2000}] is a
direct consequence of modeling the temperature conditional on (noisy)
proxy values [\citeauthor{Christiansen2010} (\citeyear{Christiansen2010,ChristiansenJC2014}),
\citet{TingleyLi2012,vonStorch04}].
Regularization may compound this problem, as with ridge regression the
smoothness of the filter factors has been shown to leak energy from the
leading SVD modes, resulting in overly damped estimates of past
temperature [\citet{SmerdonKaplan07}]. This problem may be
mitigated via TTLS [\citet{MRWA07b}], as it attempts to correct
for regression dilution by steepening the regression slope; however,
the solution is no longer guaranteed to be optimal even under broad
assumptions [\citet{CarrollRuppert96}]. Furthermore, a major
shortcoming of TTLS as currently used in climate applications is that
the truncation parameter must be specified a priori, rather than being
estimated adaptively. Given the applicability of the RegEM algorithm
for missing data problems in the paleoclimate context (e.g., surface
temperature reconstructions for the past 2000 years), we seek to
develop an imputation method that rests on a more accurate and
data-adaptive estimate of $\Sigma$ itself.

%s1.4 #&#
\subsection{Gaussian Markov random fields}\label{subsecGGM}
A GMRF is a multivariate normal model which encodes conditional
independence structure between variables [see \citet
{lauritzen,whittaker}]. More precisely, let $(X_1, \ldots, X_p)$ be a
multivariate random vector with \emph{inverse covariance matrix} (or
\emph{precision matrix}) $\Omega= (\omega_{ij}) = \Sigma^{-1}$. The
\emph{partial correlation coefficient} between $X_i$ and $X_j$ given
the rest of the variables, denoted by $\rho_{ij \mid\mathrm{rest}}$,
can be obtained from the inverse covariance matrix [see \citet
{whittaker}, Corollary 5.8.2], and is given as follows:
%
%e1.5 #&#
%
\begin{equation}
\rho_{ij \mid\mathrm{rest}} = \frac{-\omega_{ij}}{\sqrt{\omega_{ii}
\omega_{jj}}}.
\end{equation}
In the case of multivariate normal data, one can show that $\rho
_{ij\mid\mathrm{rest}} = 0$ if and only if $X_i$ is independent of $X_j$
given the rest of the variables [\citet{whittaker}, Corollary
6.3.4]. The zeros in the precision matrix therefore indicate
conditional independence between the corresponding variables. The
conditional independence relations in a distribution can be
conveniently encoded using a \emph{graph}. Recall that a graph $G =
(V,E)$ is a pair of sets $V$ and $E \subseteq V \times V$, where each
element of $V$ represents a vertex of the graph and each point of $E$
is a pair of elements of $V$. We encode the conditional independence
relations by adding an edge between $i$ and $j$ if and only if $X_i$ is
not conditionally independent of $X_j$ given the rest of the variables.
The random vector $(X_1, \ldots,X_p)$ is then said to satisfy the \emph
{pairwise Markov property} with respect to the graph $G$. For details
on the \emph{pairwise}, \emph{local} and \emph{global} Markov
properties, we refer the reader to \citet{lauritzen} and
\citet{whittaker}.

Once the conditional independence structure (or graphical structure) of
a Gaussian random vector is known, this information can be used for
estimating its covariance matrix $\Sigma$. More specifically, given an
i.i.d. sample $x_1, \ldots, x_n$ of $(X_1, \ldots, X_p)$ with mean
$\overline{x} = \frac{1}{n} \sum_{i=1}^n x_i$, and a graph $G$, the
\emph{graphical maximum likelihood estimator} of $\Sigma$ can be
computed by solving
%e1.6 #&#
%
\begin{equation}
\label{eqngraphicalMLE} \hat{\Sigma}_G = \mathop{\mathop{\argmax\limits_{\Sigma= \Omega
^{-1} >
0}}}_{\omega_{ij} = 0, (i,j) \notin E}
\log\det\Omega- \tr(S \Omega),
\end{equation}
where $S$ is the sample covariance matrix of $x_1, \ldots, x_n$, given by
%e1.7 #&#
%
\begin{equation}
S = \frac{1}{n}\sum_{i=1}^n
(x_i-\overline{x}) (x_i-\overline{x})^\top,
\end{equation}
and $\log\det\Omega- \tr(S \Omega)$ is (up to a constant) the
multivariate normal profile log-likelihood function. The problem (\ref
{eqngraphicalMLE}) can be solved efficiently for up to a few thousand
variables using, for example, regression-based algorithms [see
\citet{ESL08}, Algorithm 17.1]. The resulting matrix $\hat
{\Sigma}_G$ is generally a better estimate than the widely used sample
covariance matrix, especially when the number of observations $n$ is
smaller than the number of variables $p$.

In this paper, we propose a methodology that combines graphical models
with the EM algorithm for the purpose of reconstructing past
temperature fields. In our approach, we first model the conditional
independence structure of the target field based on structural
assumptions or directly from the data. A sparse estimate of $\Sigma$
is then obtained in accordance with this graphical structure at every
step of the EM algorithm. This approach greatly reduces the number of
parameters to estimate, leads to better conditioned and more precise
estimates of $\Sigma$, and also exploits the natural conditional
independence structure of the spatial field. The regression step~(\ref
{eqnregression}) can then be performed using any regularization method
(or even no regularization at all). We call the resulting algorithm
GraphEM (see Algorithm \ref{algoGraphEM} in Appendix~\ref{appalgo};
see also Appendix~\ref{appEM} for the derivation of the GraphEM
algorithm within the EM framework).

The rest of the paper is structured as follows. In Section~\ref
{secmethod} we explore various methods to estimate the graphical
structure of the joint temperature/proxy field. We then test the
performance of GraphEM in a realistic geophysical context in
Sections~\ref{secexperiment} and \ref{secresults}. The characteristics of
the estimated conditional independence structures are then studied in
Section~\ref{seccharacteristics}. We conclude with a discussion section.

%s2 #&#
\section{Methodology}\label{secmethod}

Different methods have been proposed in the literature to discover the
conditional independence relations (or \emph{graphical structure}) of
a data set, in either the Bayesian or frequentist framework [see e.g.,
\citet
{banerjeeetal,dawidlauritzen,FriedmanHastieTibshirani07,letacmassam2007,rajaratnametal2008}].
In this work, we explore two different approaches: $\ell_1$-penalized
maximum likelihood [\citet
{banerjeeetal,FriedmanHastieTibshirani07,NIPSGuillotetal2012,NIPSQUIC}]
and neighborhood graphs.

%s2.1 #&#
\subsection{\texorpdfstring{$\ell_1$}{ell1}-penalized maximum likelihood}\label
{subsecl1penalizedmaxlike}

\begin{figure}

\includegraphics{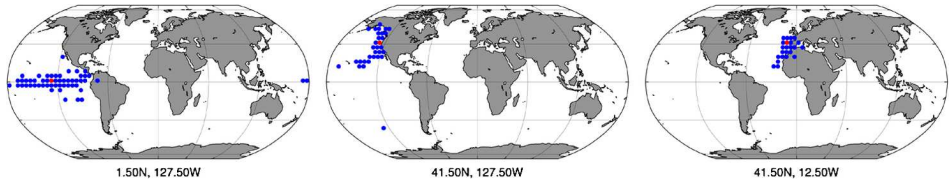}

\caption{Example of estimated graphical structure of a temperature
field (HadCRUT3v).}
\label{figgraphsrealdata}
\end{figure}

A flexible approach for obtaining a sparse estimate of the precision
matrix $\Omega$ is to maximize the normal likelihood subject to an
$\ell_1$ penalty on its norm. More specifically, the $\ell
_1$-penalized maximum likelihood problem consists of solving
%e2.1 #&#
%
\begin{equation}
\label{eqnl1likelihoodpb} \max_{\Omega> 0} l(\Omega) - \rho\llVert
\Omega\rrVert
_1,
\end{equation}
where $\Omega= \Sigma^{-1}$ denotes the precision matrix of the data,
$l(\Omega)$ is the normal log-likelihood of $\Omega$, $\rho> 0$ is a
regularization parameter, and $\llVert \Omega\rrVert _1$ is the
$1$-norm of~$\Omega$:
%e2.2 #&#
%
\begin{equation}
\llVert\Omega\rrVert_1 = \sum_{i=1}^p
\sum_{j=1}^p \llvert\omega_{ij}
\rrvert.
\end{equation}
The use of an $\ell_1$ penalty, as first introduced in the context of
the LASSO regression [\citet{Tibshirani96}], favors the
introduction of zero elements and thus leads to sparse solutions [see
\citet{ESL08}, Section~3.4.3]. At the same time, using an $\ell
_1$ penalty leads to a convex problem that can be solved efficiently
using modern methods of convex optimization. Once an estimate of
$\Omega$ is known, the associated graph can be inferred from the
pattern of zeros in $\Omega$. In this work, we employ the \emph
{graphical lasso} (glasso) algorithm of \citet
{FriedmanHastieTibshirani07} to obtain a sparse estimate of $\Omega$
by solving an $\ell_1$-penalized likelihood problem. As $\rho$
varies, the matrix $\hat{\Omega}$ displays different sparsity
patterns. When $\rho= 0$ and $n \geq p$, there is no penalty and $\hat
{\Omega}$ is equal to the maximum likelihood estimate $S^{-1}$ of
$\Omega$, where $S$ denotes the sample covariance matrix of the data
matrix. The estimate $\hat{\Omega}$ tends to a diagonal matrix as the
regularization parameter $\rho$ is increased. Problem~(\ref
{eqnl1likelihoodpb}) can also be easily modified to use a different
penalty for different parts of the matrices. Consider, for example, the
precision matrix of a temperature/proxies field. The matrix can be
organized in block form
%e2.3 #&#
%
\begin{equation}
\Omega= \pmatrix{ \Omega_{\TT} & \Omega_{\TP}
\vspace*{3pt}\cr
\Omega_{\PT} & \Omega_{\PP}},
\end{equation}
where $\Omega_{\TT}, \Omega_{\TP}$ and $\Omega_{\PP}$ are block
matrices corresponding to the temperature/temperature,
temperature/proxy and proxy/proxy parts of the matrix. Since the signal
contained in proxies is generally weaker than the temperature signal,
it may be sensible to use different penalty parameters for different
parts of the matrix when solving the $\ell_1$-penalized maximum
likelihood problem. Problem (\ref{eqnl1likelihoodpb}) can thus be
replaced by
%e2.4 #&#
%
\begin{equation}
\label{eqnl1likelihoodpbgeneral} \max_{\Omega> 0} l(\Omega) - \rho_{\TT}
\llVert\Omega_{\TT}\rrVert_1 - 2\rho_{\TP}
\llVert\Omega_{\TP}\rrVert_1 - \rho_{\PP} \llVert
\Omega_{\PP}\rrVert_1,
\end{equation}
where $\rho_{\TT}, \rho_{\TP}, \rho_{\PP} > 0$ are regularization
parameters. This problem can also be solved efficiently by using a
modified graphical lasso algorithm [see \citet
{FriedmanHastieTibshirani07}, equation (15)]. Figure~\ref
{figgraphsrealdata} displays the temperature neighbors of a few
locations for a graph estimated using (\ref
{eqnl1likelihoodpbgeneral}) (sparsity level${}=1.4\%$) on a real
temperature data set [\citet{Brohan06}], and illustrates the
potential of the $\ell_1$ method to detect real geophysical
structures. Note that the method correctly identifies anisotropic
climate features like the equatorial Pacific cold tongue (left), the
California current system (center) and east Atlantic structures related
to the subtropical gyre circulation (right).

%f3 #&#
%

In practice, choosing suitable penalty parameters in (\ref
{eqnl1likelihoodpb}) or (\ref{eqnl1likelihoodpbgeneral}) can be
difficult. A high penalty forces many zero entries in the precision
matrix, while a low penalty adds some edges that make little
geophysical sense. An optimal choice should strike a balance between
those extremes. If $\rho> \rho_{\max} := \max_{i\neq j} \llvert
S_{ij}\rrvert$,
it can be shown [see, e.g.,~\citet{Wittenetal2011}, Theorem 2]
that the resulting glasso estimate of $\Sigma$ is a diagonal matrix. A
relevant finite number of regularization parameter values can therefore
be obtained by dividing the interval between some small value $\rho
_{\min}$ and the biggest relevant value $\rho_{\max}$. In our
numerical work, we have chosen $\rho_{\min} = 0.1 \cdot\rho_{\max
}$ and have divided the interval $[\rho_{\min}, \rho_{\max}]$ into
10 values. Problem (\ref{eqnl1likelihoodpb}) can\vspace*{1pt} then be solved for
each of these penalty parameters to obtain estimates $\hat{\Omega}$
of the precision matrix $\Omega$. To each estimate corresponds a graph
based on the structure of zeros in $\hat{\Omega}$. When the dimension
of the problem to solve is small (e.g.,~in regional reconstructions) or
a single penalty parameter is used for the whole precision matrix [as
in equation (\ref{eqnl1likelihoodpb})], an optimal parameter can be
chosen using $k$-fold cross-validation. However, when a different
penalty parameter is used for each part of the precision matrix,
performing cross-validation for an array of regularization parameters
(e.g., a $10 \times10 \times10$ grid of penalty parameters) incurs a
prohibitive computational cost. A possible solution consists of
searching for a graph that is (a) dense enough to capture the salient
spatial dependences, and (b) sparse enough to make the reconstruction
possible and stable (by reducing the dimension of the problem to a size
comparable to the sample size). A~triple $(\rho_{\TT}, \rho_{\TP}, \rho
_{\PP})$ of regularization parameters with the desired sparsity can be
chosen by starting with large values of the three penalty parameters,
and progressively reducing the value of each penalty parameter until a
given target sparsity is obtained for each part of the precision
matrix. This technique requires computing the solution of problem (\ref
{eqnl1likelihoodpbgeneral}) at only a few points of the grid. This
sparsity approach is implemented in our proposed version of GraphEM,
and is compared to the neighborhood approach described below in
Section~\ref{secresults}. In this paper, we have chosen fixed sparsity levels
when performing large reconstruction ensembles, after verifying via
targeted experiments that the specified sparsity levels were close to
those deemed optimal by $5$-fold cross-validation.

%s2.2 #&#
\subsection{Neighborhood graphs}\label{subsecneighborhoodgraphs}
\begin{figure}[t]%

\includegraphics{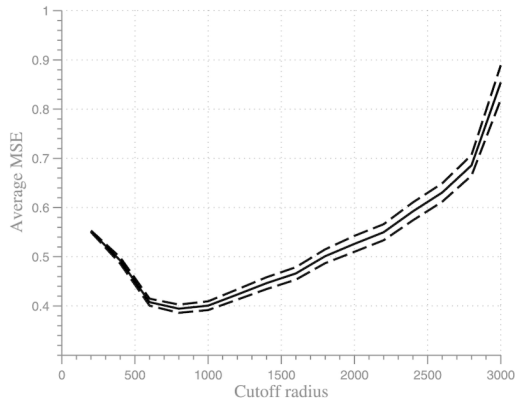}

\caption{Cross-validation scores for choosing a neighborhood graph
radius (5-fold cross-validation).}
\label{figcvneigh}
\end{figure}

Since temperature variations at a given point are to a large extent
explained by temperature of surrounding points, it is natural to use a
\emph{neighborhood graph} (i.e., a graph where two vertices are
connected if and only if they are within a specified radius $R$) to
approximate the true graphical structure of the joint temperature/proxy
field; see, for example,~\citet{Cooketal1999} where a similar
assumption was made. The radius can be either specified or chosen from
the data. As we illustrate in Figure~\ref{figcvneigh},
%f4 #&#
%
the choice of
an optimal radius can be made by performing cross-validation over the
instrumental period and choosing the radius that minimizes the MSE of
the reconstructed values. Besides this natural and meaningful Markov
random field structure in spatial temperature fields, a~neighborhood
graph approach has the distinct advantage that the underlying graph
does not have to be estimated from sample-deficient high-dimensional
data, and that the procedure does not require solving computationally
intensive optimization problems. Dimensionality reduction is achieved
with great ease and at the same time has an intuitive geophysical
interpretation; sparsity is entirely governed by the neighborhood
radius $R$. On the other hand, neighborhood graphs are less flexible
and cannot model in an adaptive way (1) conditional independence
relations resulting from anisotropic structures present in the data
(such as land/ocean boundaries, mountain ranges, atmospheric flow
patterns, etc.), and (2) long range dependencies that arise due to
teleconnections. However, when the noise level is too high, a simple
model such as a neighborhood graph may be preferable to the $\ell
_1$-penalized covariance estimation method.

%f5 #&#
%
\begin{figure}

\includegraphics{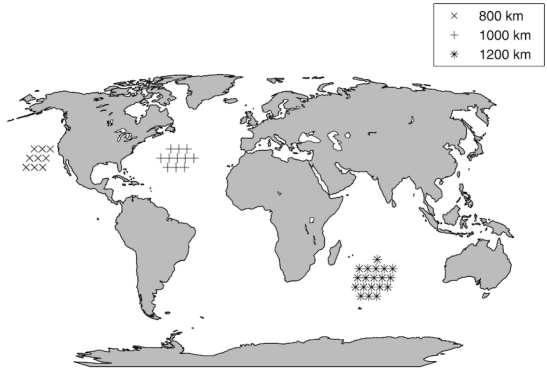}

\caption{Illustration of neighborhood graphs.}
\label{figneighgraphrealtemp}
\end{figure}

As an illustration, Figure~\ref{figneighgraphrealtemp} displays
the neighborhood of size 800 km, 1000 km and 1200 km at different
locations with the same latitude on a $5^\circ\times5^\circ$ grid.
The average number of neighbors (and their standard deviation) are 8.42
(2.08), 10.24 (3.37) and 16.61 (4.92), respectively.

We also consider sparser variants of the neighborhood graph model based
on the structure of the paleoclimate reconstruction problem. First,
since climate is the signal shared between proxies, it is natural to
assume that the proxies are independent of each other conditional on
the temperature data (i.e.,~to assume that $\Omega_{\PP}$ is diagonal).
We thus explore a simpler model where the temperature/temperature (TT)
and the temperature/proxy (TP) parts of the graph are constructed as
above with a neighborhood graph, but where $\Omega_{\PP}$ is diagonal.
Further, since temperature proxies are reflective of local temperature
only, it is natural to impose a local structure in $\Omega_{\PT}$ as
well (i.e., $\Omega_{\PT} = 0$ except for the columns corresponding to
each proxy's closest temperature grid point). Finally, given that the
optimal neighborhood graphs chosen by cross-validation tend to feature
only the immediate neighbors of each temperature gridpoint, it is
natural to impose such constraints on the TT part of the graph {a
priori}. Note that such a model is equivalent to a spatial
conditionally auto-regressive (CAR) model [\citet{besag74}]. The
variants considered in the paper are summarized in Table~\ref{tablegraphs}, and their performance in modeling the conditional
independence structure of the temperature/proxy field is studied in
Section~\ref{sec4}.\vadjust{\goodbreak}
%
%t1 #&#
%
\begin{table}%[H]
\tabcolsep=0pt
\caption{Neighborhood graph variants for the joint temperature/proxy field}
\label{tablegraphs}
\begin{tabular*}{\tablewidth}{@{\extracolsep{\fill}}@{}lccc@{}}
\hline
\textbf{Name} & \textbf{TT} & \textbf{TP} & \textbf{PP}\\
\hline
Neigh & Neighborhood & Neighborhood & Neighborhood \\
$\mathrm{Ind}_{\mathrm{PP}}$ & Neighborhood & Neighborhood & Diagonal\\
$\mathrm{CAR}_{\mathrm{TP}}$ & Neighborhood & CAR & Diagonal \\
$\mathrm{CAR}_{\mathrm{TT}}$ & CAR & Neighborhood & Diagonal \\
$\mathrm{CAR}_{\mathrm{TT}+\mathrm{TP}}$ & CAR & CAR & Diagonal\\
\hline
\end{tabular*}
\end{table}

%s3 #&#
\section{Validation via pseudoproxy experiments}\label{secexperiment}
%s3.1 #&#
\subsection{Background}
In the climate literature, pseudoproxy experiments have become the
method of choice to objectively evaluate the performance of CFR
techniques against a geophysically-relevant target [see \citet
{SmerdonWIRES12} for a recent review]. This target temperature field is
often the output of coupled general circulation model (GCM) simulations
for the past 1000 years or so, sampled at a fixed spatiotemporal
resolution. Although GCM-simulated temperature fields do not exactly
match the characteristics of observed temperature fields, they are
generated in accordance with physical laws embedded in such models, and
thus provide a controlled, realistic framework to test reconstruction methods.

In practice, a pseudoproxy is obtained by adding noise to a
GCM-simulated temperature field at locations where proxy observations
are available in the real world. Because such observations are sparse,
the pseudoproxy network therefore comprises a small collection of time
series. Given only knowledge of the temperature field over a 150-year
calibration interval, the CFR method is then used to backcast a
thousand-year long temperature field based on this relatively small
sample of noisy temperature time series. Given a simulated temperature
field $T(l,t)$ at location $l$ and time $t$ (standardized to have mean
$0$ and variance $1$ over time) from a GCM model, the pseudoproxies
$P(l,t)$ are constructed as follows:
%e3.1 #&#
%
\begin{equation}
P(l,t) = T(l,t) + \frac{1}{\mathrm{SNR}} \cdot\xi(l,t),
\end{equation}
where $\xi(l,t)$ are independent realizations of a Gaussian white
noise process, and the (scalar) signal-to-noise ratio $\mathrm{SNR}$
controls the amount of noise in the pseudoproxy. Although pseudoproxies
constitute an oversimplification of reality, they have been used
extensively in the climate literature [\citet
{Annan2012,Bradley1996,BoChris09,Li2012,MannRutherford02,SmerdonGRL11,SmerdonWIRES12,TingleyHuybers10b}]
to provide a numerical laboratory to test the performance of CFR methods.

%f6 #&#
%
\begin{figure}

\includegraphics{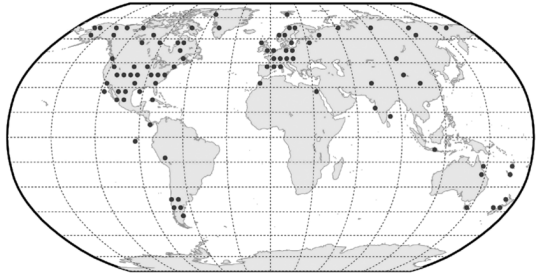}

\caption{Geographic location of the pseudoproxies in the MBH98 database.}
\label{figMBH98}
\end{figure}

In our simulations, we used the NCAR CSM 1.4 model experiment
[\citet{Ammann2007}], which simulates the climate of the last
millennium (850--1980 AD) on a $5^\circ\times5^\circ$ grid. As per
previous work [\citet
{LietalJASA2010,MRWA07,SmerdonGRL11,WangCP2014}], the locations of the
pseudoproxies were chosen in accordance with \citet{MBH98} (MBH98,
Figure~\ref{figMBH98}) and the value of $\mathrm{SNR}$ has been
fixed to $0.5$. Other SNR values have also been investigated but, for
the sake of brevity, are not presented here.
The last 150 years of data have been used as a calibration
period, and the remaining 981 years of temperature data have been
reconstructed using GraphEM. As a benchmark, we follow recent work
[\citet{SteigerJC2014,TingleyHuybers10b}] and use RegEM-TTLS,
which was widely used in high-profile climate reconstructions
[\citeauthor{Mann08a} (\citeyear{Mann08a,Mann09})].

%s3.2 #&#
\subsection{Performance metrics}
Various metrics have been used in the literature to measure the quality
of CFR methods and reconstructed temperature fields [\citet
{BurgerCP07,CookBriffaJones94}]. Let $T(l,t)$ denote the temperature at
a location $l$ and at time $t$, and denote by $\hat{T}(l,t)$ a
reconstruction of $T(l,t)$. The \emph{mean squared error} (MSE)
measures the mean difference between the two fields at a given location $l$:
%e3.2 #&#
%
\begin{equation}
\MSE(\hat{T}) (l) = \frac{1}{N}\sum_t
\bigl(T(l,t)-\hat{T}(l,t)\bigr)^2,
\end{equation}
where $N$ is the number of time points. To measure the improvement made
by our proposed graphical method, we define the \textit{relative MSE
difference} at a location $l$ by
\[
\mbox{relative MSE difference $(l)$} = \frac{\mathrm{MSE}_\mathrm
{RegEM\mbox{-}TTLS}(l)-\mathrm{MSE}_\mathrm{GraphEM}(l)}{\mathrm
{MSE}_\mathrm{RegEM\mbox{-}TTLS}(l)}.
\]
Although a small MSE indicates a good reconstruction, it is not
immediately clear how small the MSE has to be for the reconstruction to
be considered a ``good reconstruction.'' A useful approach is to
compare the MSE of a given reconstruction to that of a reconstruction
that is equal to a constant value over time (a ``constant
reconstruction''). The \emph{reduction of error} (RE) compares the MSE
of a given reconstruction to a constant reconstruction equal to the
mean temperature of the field $\overline{T}_c(l)$ over the calibration period:
%e3.3 #&#
%
\begin{equation}
\mathrm{RE}(l) = 1 - \frac{\MSE(\hat{T})(l)}{\MSE(\overline{T}_c)(l)}.
\end{equation}
Similarly, the \emph{coefficient of efficiency} (CE) compares the MSE
of the reconstruction to a constant reconstruction equal to the mean of
the temperature field $\overline{T}_v(l)$ over the validation interval:
%e3.4 #&#
%
\begin{equation}
\mathrm{CE}(l) = 1 - \frac{\MSE(\hat{T})(l)}{\MSE(\overline{T}_v)(l)}.
\end{equation}
Finally, the bias at point $l$ is the difference between $\hat
{T}(l,\cdot)$ and $T(l,\cdot)$ averaged over time. A perfect
reconstruction would have a MSE of $0$, a CE and a RE of $1$ and a bias
of $0$. The closer to those values, the better the reconstruction.

%s4 #&#
\section{Results} \label{secresults}\label{sec4}
In order to test the performance and the sensitivity of\break GraphEM to
reconstruct temperature over the whole globe, we performed $50$
reconstructions, each corresponding to a different noise realization
$\xi(l,t)$. The performance of GraphEM is then compared to the
performance of RegEM-TTLS. The truncation parameter was set to 5, but
the results show little sensitivity to this choice.

To study the performance of GraphEM, reconstructions were performed
using both the neighborhood graph methods and the $\ell_1$ method
(Section~\ref{secmethod}). For illustration purposes, in
Sections~\ref{secresultsspatial} and \ref{secglobalmean}, we
present detailed results for the neighborhood graph method with a
cutoff radius of $800$ km, as suggested by \mbox{cross-}validation (see
Figure~\ref{figcvneigh}). Verification statistics for other cutoff radii,
for the neighborhood graph variants and for the $\ell_1$ method are
also provided in Tables~\ref{tableglobstats0p5compare}~and~\ref{tableglobstatmean0p5}.

%s4.1 #&#
\subsection{Spatial reconstructions}\label{secresultsspatial}
\begin{figure}[t]

\includegraphics{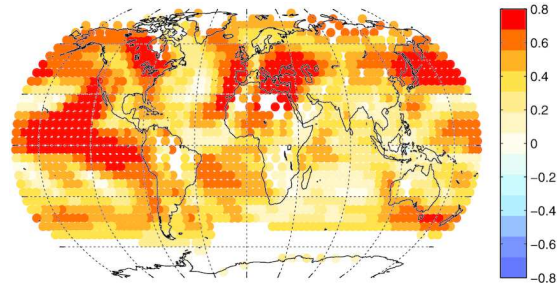}

\caption{Relative MSE improvement for SNR${}={}$0.5 (neighborhood graph,
cutoff radius${}={}$800 km).}
\label{figgloberr}
\end{figure}

We begin by studying the performance of GraphEM in space. Figure~\ref
{figgloberr} displays the average relative MSE improvement for the
$50$ reconstructions, and shows that the improvement can be substantial
when using GraphEM. The improvement is positive for almost every
location. The average improvement is about 43\%, whereas improvements
as large as 80\% are recorded in certain regions. Figure~\ref
{figgloberr} also provides some compelling evidence that the
magnitude of the percentage improvement appears to be even greater at
some locations that are distant from proxy sites. In particular, vast
swathes of the entire central and northern Pacific stretching from East
Asia to North and central America display significantly higher
improvements in MSE. The same appears to be true for parts of the
southern Atlantic. This is remarkable given the high degree of locality
of the chosen graph. Hence, a local graph does not translate into
short-range correlations; on the contrary, it can actually improve the
representation of long-range dependencies. Improvements over the Indian
ocean, however, tend to be modest perhaps because of the paucity of data.

%f7 #&#
%

%f8 #&#
%
\begin{figure}[b]

\includegraphics{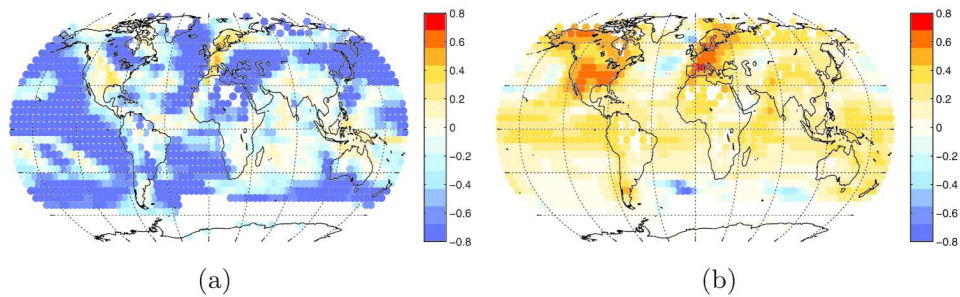}

\caption{CE map for the \textup{(a)}~RegEM-TTLS and \textup{(b)}~GraphEM reconstructions for
SNR${}= 0.5$ (neighborhood graph, cutoff radius${}={}$800
km).}\label{figglobCEttls}\label{figglobCEggm}
\end{figure}

Figure~\ref{figglobCEttls}(a)~and~(b) display the CE statistics
(averaged over the $50$ noise realizations) for RegEM-TTLS and GraphEM,
respectively. Again, in many regions, GraphEM leads to substantial
improvements, particularly where the skill was very poor with
RegEM-TTLS. The different precision metrics averaged over space (for
the unsmoothed reconstruction) are presented in Table~\ref
{tableglobstats0p5compare} along with their standard deviation
computed using the $50$ reconstructions. This table confirms once more
that GraphEM performs better spatially and is more stable than RegEM-TTLS.

%t2 #&#
%
\begin{table}
\tabcolsep=0pt
\caption{Mean (and standard deviation) of the performance metrics
averaged over space for the global reconstructions}\label{tableglobstats0p5compare}
\begin{tabular*}{\tablewidth}{@{\extracolsep{\fill}}@{}lcccc@{}}
\hline
\textbf{Method} & \textbf{MSE} & \textbf{RE} & \textbf{CE} & \textbf{Bias} \\
\hline
$\ell_1$ method\\
\quad GraphEM ($0.3$\% target sparsity) & 0.44 (0.01) & 0.33 (0.01) & 0.11 (0.02) & 0.09 (0.01) \\
\quad GraphEM ($0.5$\% target sparsity) & 0.42 (0.01) & 0.36 (0.01) & 0.15 (0.01) & 0.08 (0.01) \\
\quad GraphEM ($0.7$\% target sparsity) & 0.41 (0.01) & 0.36 (0.01) & 0.16 (0.01) & 0.08 (0.01) \\
\quad GraphEM ($0.9$\% target sparsity) & 0.41 (0.01) & 0.36 (0.01) & 0.15 (0.01) & 0.08 (0.01)
\\[2.5pt]
Neigh\\
\quad GraphEM ($600$ km radius) & 0.42 (0.01) & 0.35 (0.01) & 0.14 (0.01) & 0.06 (0.01) \\
\quad GraphEM ($800$ km radius) & 0.39 (0.01) & 0.39 (0.01) & 0.19 (0.01) & 0.06 (0.01) \\
\quad GraphEM ($1000$ km radius) & 0.40 (0.01) & 0.38 (0.01) & 0.18 (0.01) & 0.06 (0.01) \\
\quad GraphEM ($1200$ km radius) & 0.41 (0.01) & 0.36 (0.01) & 0.16 (0.01) & 0.06 (0.01)
\\[2.5pt]
$\mathrm{Ind}_{\mathrm{PP}}$\\
\quad GraphEM ($600$ km radius) & 0.42 (0.01) & 0.35 (0.01) & 0.13 (0.01) & 0.06 (0.01) \\
\quad GraphEM ($800$ km radius) & 0.39 (0.01) & 0.39 (0.01) & 0.19 (0.01) & 0.06 (0.01) \\
\quad GraphEM ($1000$ km radius) & 0.40 (0.01) & 0.38 (0.01) & 0.19 (0.01) & 0.06 (0.01) \\
\quad GraphEM ($1200$ km radius) & 0.41 (0.01) & 0.37 (0.01) & 0.16 (0.01) & 0.06 (0.01)
\\[2.5pt]
$\mathrm{CAR}_{\mathrm{TP}}$\\
\quad GraphEM ($600$ km radius) & 0.42 (0.01) & 0.35 (0.01) & 0.14 (0.01) & 0.06 (0.01) \\
\quad GraphEM ($800$ km radius) & 0.39 (0.01) & 0.39 (0.01) & 0.19 (0.01) & 0.06 (0.01) \\
\quad GraphEM ($1000$ km radius) & 0.39 (0.01) & 0.39 (0.01) & 0.19 (0.01) & 0.06 (0.01) \\
\quad GraphEM ($1200$ km radius) & 0.40 (0.01) & 0.38 (0.01) & 0.18 (0.01) & 0.06 (0.01)
\\[2.5pt]
$\mathrm{CAR}_{\mathrm{TT}}$\\
\quad GraphEM ($600$ km radius) & 0.39 (0.01) & 0.39 (0.01) & 0.20 (0.01) & 0.06 (0.01) \\
\quad GraphEM ($800$ km radius) & 0.39 (0.01) & 0.39 (0.01) & 0.19 (0.01) & 0.06 (0.01) \\
\quad GraphEM ($1000$ km radius) & 0.40 (0.01) & 0.39 (0.01) & 0.19 (0.01) & 0.06 (0.01) \\
\quad GraphEM ($1200$ km radius) & 0.40 (0.01) & 0.38 (0.01) & 0.18 (0.01) & 0.06 (0.01)
\\[2.5pt]
$\mathrm{CAR}_{\mathrm{TT}+\mathrm{TP}}$\\
\quad GraphEM ($600$ km radius) & 0.39 (0.01) & 0.39 (0.01) & 0.19 (0.01) & 0.06 (0.01) \\
\quad GraphEM ($800$ km radius) & 0.39 (0.01) & 0.39 (0.01) & 0.19 (0.01) & 0.06 (0.01) \\
\quad GraphEM ($1000$ km radius) & 0.39 (0.01) & 0.39 (0.01) & 0.19 (0.01) & 0.06 (0.01) \\
\quad GraphEM ($1200$ km radius) & 0.39 (0.01) & 0.39 (0.01) & 0.19 (0.01) & 0.06 (0.01)
\\[2.5pt]
RegEM-TTLS\\
\quad RegEM-TTLS & 0.84 (0.10) & $-$0.24 (0.14) & $-$0.61 (0.19) & 0.01 (0.02) \\
\hline
\end{tabular*}\vspace*{-6pt}
\end{table}

Although the results presented in Table~\ref{tableglobstats0p5compare}
are quite similar for the different
GraphEM methods, the neighborhood graphs seem to perform slightly
better than the $\ell_1$ method. They could therefore be useful in
noisy cases for which discovering the structure of the field from the
data is difficult. Another advantage of the neighborhood method is that
the cutoff radius is easy to choose by cross-validation.\vadjust{\goodbreak} In comparison,
choosing appropriate regularization parameters to use with the $\ell
_1$ method is computationally intensive.

%f9 #&#
%
\begin{figure}[b]%

\includegraphics{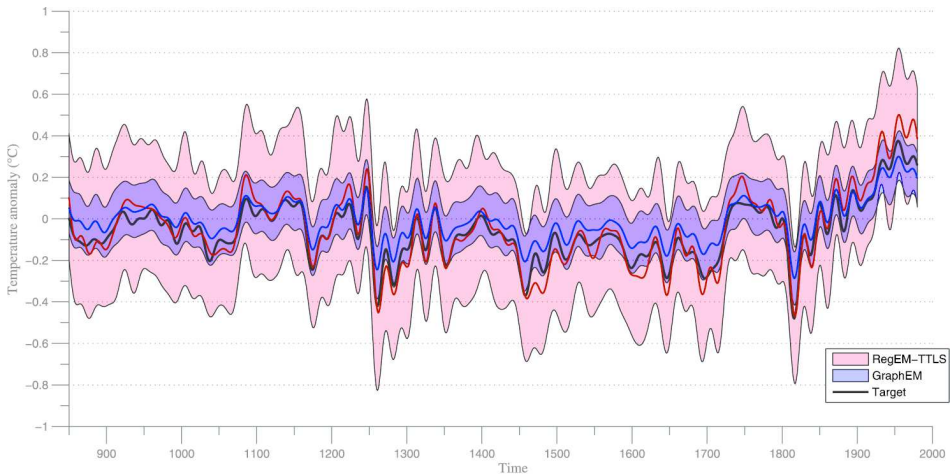}

\caption{Global spatial averages (multiple noise realizations, 95\%
deviation interval) for SNR${} = 0.5$ (neighborhood graph,
cutoff radius${}={}$800 km). The thick lines represent the median of each
ensemble.}
\label{figglobmean}
\end{figure}

We also observe that the four neighborhood graph variants produce very
similar results. In particular, the simplest graph $\mathrm{CAR}_{\TT}$
performs quite well, underlying the importance of locality in the
temperature/proxy field. In the pseudo proxy experiment, the better
validation metrics also correspond to the models that best reflect the
data generating mechanism, that is, the models where $\Omega_{\PP}$ is
diagonal. Climate fields found in nature may display a more complex
structure, but to the extent that it can be reasonably approximated by
a neighborhood graph, our results suggest that GraphEM could produce
very skillful reconstructions.

%
%t3 #&#
%
\begin{table}[b]
\tabcolsep=0pt
\caption{Mean (and standard deviation) of the performance metrics for
the spatial average reconstructions}\label{tableglobstatmean0p5}
\begin{tabular*}{\tablewidth}{@{\extracolsep{\fill}}@{}lcccc@{}}
\hline
\textbf{Method} & \textbf{MSE} & \textbf{RE} & \textbf{CE} & \textbf{Bias}\\
\hline
$\ell_1$ method & & & &  \\
\quad GraphEM ($0.3\%$ target sparsity) & 0.12 (0.01) & 0.75 (0.02) & 0.25 (0.08) & 0.09 (0.01) \\
\quad GraphEM ($0.5\%$ target sparsity) & 0.11 (0.01) & 0.79 (0.02) & 0.36 (0.05) & 0.08 (0.01) \\
\quad GraphEM ($0.7\%$ target sparsity) & 0.11 (0.01) & 0.79 (0.02) & 0.37 (0.05) & 0.08 (0.01) \\
\quad GraphEM ($0.9\%$ target sparsity) & 0.11 (0.01) & 0.79 (0.02) & 0.36 (0.05) & 0.08 (0.01)
\\[2.5pt]
Neigh \\
\quad GraphEM ($600$ km) & 0.12 (0.01) & 0.82 (0.01) & 0.46 (0.04) & 0.06 (0.01) \\
\quad GraphEM ($800$ km) & 0.11 (0.01) & 0.83 (0.01) & 0.50 (0.04) & 0.06 (0.01) \\
\quad GraphEM ($1000$ km) & 0.10 (0.01) & 0.83 (0.01) & 0.50 (0.04) & 0.06 (0.01) \\
\quad GraphEM ($1200$ km) & 0.10 (0.01) & 0.83 (0.01) & 0.48 (0.04) & 0.06 (0.01)
\\[2.5pt]
$\mathrm{Ind}_{\mathrm{PP}}$ & &  & & \\
\quad GraphEM ($600$ km radius) & 0.12 (0.01) & 0.82 (0.01) & 0.46 (0.04) & 0.06 (0.01) \\
\quad GraphEM ($800$ km radius) & 0.11 (0.01) & 0.83 (0.01) & 0.50 (0.04) & 0.06 (0.01) \\
\quad GraphEM ($1000$ km radius) & 0.11 (0.01) & 0.83 (0.01) & 0.50 (0.04) & 0.06 (0.01) \\
\quad GraphEM ($1200$ km radius) & 0.11 (0.01) & 0.83 (0.01) & 0.48 (0.04) & 0.06 (0.01)
\\[2.5pt]
$\mathrm{CAR}_{\mathrm{TP}}$ & & & & \\
\quad GraphEM ($600$ km radius) & 0.12 (0.01) & 0.83 (0.01) & 0.47 (0.04) & 0.06 (0.01) \\
\quad GraphEM ($800$ km radius) & 0.11 (0.01) & 0.83 (0.01) & 0.49 (0.04) & 0.06 (0.01) \\
\quad GraphEM ($1000$ km radius) & 0.11 (0.01) & 0.83 (0.01) & 0.50 (0.04) & 0.06 (0.01) \\
\quad GraphEM ($1200$ km radius) & 0.11 (0.01) & 0.83 (0.01) & 0.49 (0.04) & 0.06 (0.01)
\\[2.5pt]
$\mathrm{CAR}_{\mathrm{TT}}$ &  & & & \\
\quad GraphEM ($600$ km radius) & 0.11 (0.01) & 0.83 (0.01) & 0.49 (0.04) & 0.06 (0.01) \\
\quad GraphEM ($800$ km radius) & 0.11 (0.01) & 0.83 (0.01) & 0.49 (0.04) & 0.06 (0.01) \\
\quad GraphEM ($1000$ km radius) & 0.11 (0.01) & 0.83 (0.01) & 0.49 (0.04) & 0.06 (0.01) \\
\quad GraphEM ($1200$ km radius) & 0.11 (0.01) & 0.84 (0.01) & 0.50 (0.04) & 0.06 (0.01)
\\[2.5pt]
$\mathrm{CAR}_{\mathrm{TT}+\mathrm{TP}}$ & & & & \\
\quad GraphEM ($600$ km radius) & 0.11 (0.01) & 0.83 (0.01) & 0.47 (0.04) & 0.06 (0.01) \\
\quad GraphEM ($800$ km radius) & 0.11 (0.01) & 0.83 (0.01) & 0.47 (0.04) & 0.06 (0.01) \\
\quad GraphEM ($1000$ km radius) & 0.11 (0.01) & 0.83 (0.01) & 0.47 (0.04) & 0.06 (0.01) \\
\quad GraphEM ($1200$ km radius) & 0.11 (0.01) & 0.83 (0.01) & 0.47 (0.04) & 0.06 (0.01)
\\[2.5pt]
RegEM-TTLS \\
\quad RegEM-TTLS & 0.15 (0.03) & 0.63 (0.18) & $-$0.12 (0.56) & 0.01 (0.02)\\
\hline
\end{tabular*}\vspace*{-6pt}
\end{table}

The results also demonstrate that a larger graph (e.g.,~neighborhood
1200 km vs CAR) can still lead to a very good reconstruction. This is
to be expected since an edge between two vertices does not prohibit the
corresponding entry in $\Omega$ from being very small. Thus, a graph
containing a certain number of spurious edges (such as the graphs
obtained from the $\ell_1$ method) may still perform well, which means
that results are broadly insensitive to the graph density. Finally, we
note that although the $\ell_1$ method performs slightly worse in our
experiments, it has the potential to detect real geophysical
structures, and could lead to improvements when working with data sets
with a stronger signal.

%s4.2 #&#
\subsection{Spatial average}\label{secglobalmean}
The spatial reconstructions given by RegEM-TTLS and GraphEM can also be
averaged over space to obtain (area-weighted) spatial averages.
Figure~\ref{figglobmean} displays a $95\%$ deviation band
(constructed using the $50$ reconstructions) for the mean temperature
series reconstructed with RegEM-TTLS and GraphEM. The instrumental
period is also reconstructed via the pseudoproxies\vadjust{\goodbreak} using the estimated
mean and covariance matrix obtained from GraphEM. The uncertainty bands
have been obtained by computing the (weighted) average temperature at
each time for each reconstruction, and then constructing a confidence
interval containing $95\%$ of the 50 simulated values. A 20 year
low-pass filter has been applied after computing the quantiles for
illustration and interpretation purposes. The mean width of the
deviation interval for GraphEM and RegEM-TTLS are $0.25$ and $0.66$,
respectively. The associated reconstruction statistics are provided in
Table~\ref{tableglobstatmean0p5}. Note that the CE scores for
GraphEM are significantly larger\vadjust{\goodbreak} than the corresponding scores for
RegEM-TTLS. Moreover, the standard deviations of the CE scores are
significantly smaller for GraphEM. The results thus demonstrate that
GraphEM can also be useful for reconstructing indices such as the mean
temperature, with better risk properties than RegEM-TTLS.

%s4.3 #&#
\subsection{Uncertainty quantification}\label{secboot}
\begin{figure}[b]%

\includegraphics{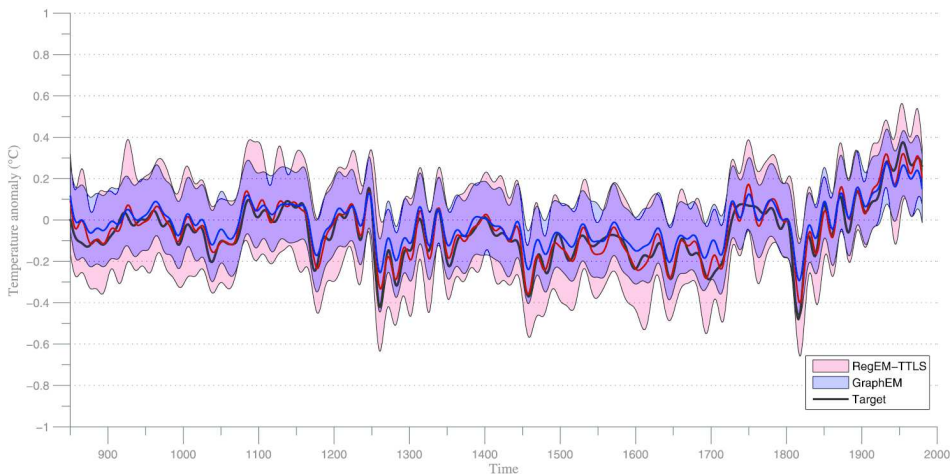}

\caption{Inflated spatial average uncertainty estimated by
nonparametric bootstrap, blocksize${}= 2$ (neighborhood graph, cutoff
radius${}={}$800 km). The thick lines represent the median of each ensemble.}\label{figmeanglobalbootstrap}
\end{figure}

Section~\ref{secresultsspatial} demonstrates the ability of GraphEM
to reduce the uncertainties in paleoclimate reconstructions via an
ensemble of pseudoproxies. In practice, it is necessary to obtain an
estimate of the uncertainties internally [see, e.g., \citet
{LietalJASA2010}]. We therefore produce prediction intervals for both
RegEM and GraphEM using a nonparametric block bootstrap method
[\citet{liubootstrap}]. The technique is described in Appendix~\ref
{appbootstrap}, and is illustrated for the global reconstruction
of Section~\ref{secresultsspatial}. Using the reconstruction $\hat
{X}_1, \ldots, \hat{X}_N$ provided by the nonparametric bootstrap, we
estimate a $95\%$ prediction interval for each reconstructed mean by
computing the $2.5$th and $97.5$th percentiles of the empirical
distribution. The mean width of the uncertainty bands for GraphEM and
RegEM-TTLS are $0.35$ and $0.45$, respectively. Comparing Figures~\ref
{figglobmean} and \ref{figmeanglobalbootstrap}, we observe that
the uncertainties of GraphEM seem slightly overestimated, whereas the
uncertainties of RegEM-TTLS seem underestimated by the bootstrap.

%f10 #&#
%

The coverage rates over the validation period for GraphEM and
RegEM-TTLS are, respectively, 92.3\% and 91.4\%. The coverage rates of
our method thus appear reasonable. Two natural techniques can be used
if a given coverage rate needs to be obtained: (1) modify the band
width to obtain the right coverage, or (2) inflate the variance of the
reconstructed values in the bootstrap [see \citet
{jansonrajaratnam,LietalJASA2010} for details]. Recall that in our
reconstructions, the instrumental period is also reconstructed using
the pseudoproxies. The reconstructed values over the instrumental
period can thus provide guidance about how much to inflate the
uncertainty bands to obtain a given coverage rate. In Figure~\ref
{figmeanglobalbootstrap}, the coverage rates over the instrumental
period for GraphEM and RegEM-TTLS are 91.3\% and 90.7\%, respectively.
In order to obtain a coverage rate of, say, 95\% over the instrumental
period, the GraphEM and RegEM-TTLS bands must be inflated by a factor
of $1.15$ and $1.42$, respectively. Inflating the bands by these factors
yields coverage rates of 94.2\% and 97.2\% on the validation period,
respectively. Inflation factors can also be computed in a more
principled way by using $k$-fold cross-validation over the instrumental
period. In our simulations, we split the instrumental period into $5$
blocks and used the bootstrap to reconstruct each block using the other
$4$ blocks. In each case, an inflation factor can be computed so that
the uncertainty bands cover $95\%$ of the targeted mean over the
instrumental period. Using this technique, we obtained an average
inflation factor of $1.10$ with GraphEM (similar to the inflation
factor obtained without cross-validation).

%s5 #&#
\section{Characteristics of paleoclimate Markov random fields}\label{seccharacteristics}

Our results demonstrate that the GraphEM approach produces substantial
improvements in comparison to RegEM-TTLS almost uniformly over space.
This section examines the characteristics of paleoclimatic Markov
random fields. More precisely, we study the properties of the joint
temperature/proxy graph estimated using the $\ell_1$ method, with the
goal of understanding (a) whether the GraphEM approach is indeed
achieving its original aim of parameter reduction, and (b) what are the
important features of estimated temperature/proxy fields. In
particular, we examine the difference between the graphical structures
estimated from the data using the $\ell_1$ method (Section~\ref
{subsecl1penalizedmaxlike}) and the neighborhood structures
described in Section~\ref{subsecneighborhoodgraphs}.

%f11 #&#
%
\begin{figure}
\includegraphics{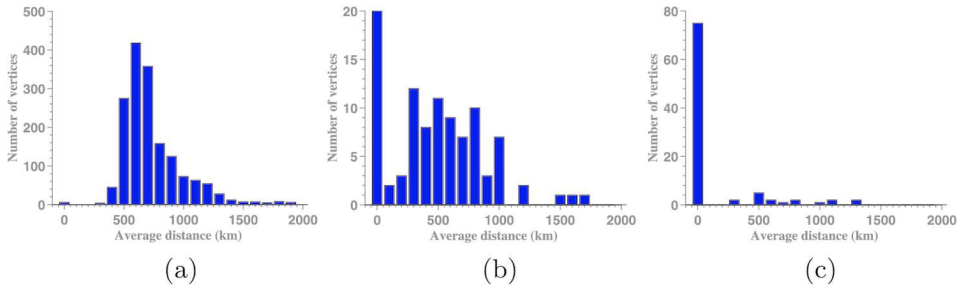}

\caption{Distribution of the average distance to each vertex in the
different part of the graph (sparsity level${}={}$0.5\%, SNR${}= 0.5$).
\textup{(a)}~TT,
\textup{(b)}~TP,
\textup{(c)}~PP.}\label{figdistancedistribution}
\end{figure}

%f12 #&#
%
\begin{figure}[b]%

\includegraphics{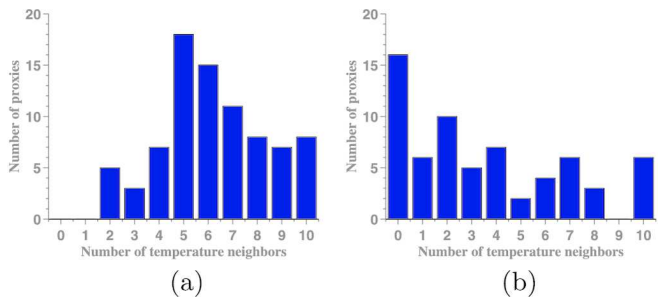}

\caption{Distribution of the number of temperature neighbors of proxy
points (sparsity level${}={}$0.5).
\textup{(a)}~SNR${} = \infty$,
\textup{(b)}~SNR${} = 0.5$.}\label{temphistdegreeprox}\label{figtemphistdegreeproxinf}\label{figtemphistdegreeprox0p5}
\end{figure}

We first illustrate the achieved parameter reduction when the graph is
estimated from the data. Figure~\ref{figdistancedistribution}
displays the distribution of the average distance from each vertex to
its neighbors in the TT, TP and PP part of the temperature/proxy graph
estimated with the $\ell_1$ method with a sparsity level of $0.5\%$ in
each part of the precision matrix. We observe that each point is
generally only connected to geographically close neighbors, although
the graph can display some far away connections (which may or may not
represent geophysical relations). The average number of neighbors in
the TT, TP and PP parts of the graph are 10.5, 9.4 and 0.42,
respectively. The graph therefore displays a neighborhood structure in
the TT and TP part of the graph, with a cutoff radius of roughly 800
km. Note also that the absence of connections in the PP part of the
graph suggests that proxies are conditionally independent given the
temperature data, and that the estimated graph is very similar to the 2
families of graphs described at the end of Section~\ref
{subsecneighborhoodgraphs}. The main message is that the number of
neighbors is relatively few compared to one that would be present with
a full precision matrix, and that locality seems to be an important
characteristic of paleoclimate Markov random fields.

Figure~\ref{temphistdegreeprox} displays the distribution of the
number of temperature neighbors of each proxy (in the graph estimated
with the $\ell_1$ method) when no noise has been added to the
temperature time series when generating pseudoproxies ($\mathrm{SNR} =
\infty$), as compared to the typical noise case that has been studied
thus far ($\mathrm{SNR} = 0.5$). Both graphs have been obtained using
the $\ell_1$ method with a sparsity level of $0.5\%$. This comparison
shows that many proxies do not have any temperature neighbors in the
$\mathrm{SNR} = 0.5$ case. In comparison, a relation between each
proxy and some temperature locations has been detected in the $\mathrm
{SNR} = \infty$ case. Detecting temperature/proxy relations from the
data can thus be an issue when the level of noise is high. The
potential for the $\ell_1$ method to detect spurious relations in the
presence of noise is also to be expected [\citet{banerjeeetal}].
This problem may be mitigated by adding further constraints on the
estimated graph. Neighborhood graphs offer a natural solution and
provide a good graphical structure independently of the level of noise.

%f13 #&#
%
\begin{figure}

\includegraphics{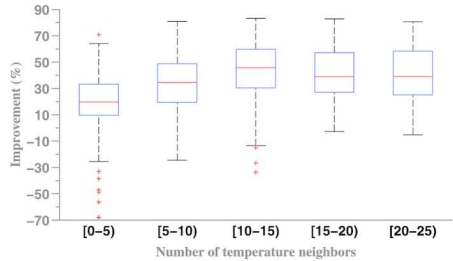}

\caption{Boxplot of the \% improvement as a function of the number of
temperature neighbors (sparsity level${}={}$0.5\%, SNR${} = 0.5$).}
\label{figboxplotimprovdegree}
\end{figure}

We now examine how sparsity translates to improvements in paleoclimate
reconstructions. Figure~\ref{figboxplotimprovdegree} displays the
improvements given by GraphEM (as compared to RegEM-TTLS) at different
temperature points vs. their connectivity (number of temperature
neighbors) in the corresponding graph.

The figure indicates that percentage improvement is smaller for
temperature points with very few neighbors. The improvement is maximal
when the number of neighbors roughly corresponds to the number of
immediate geographical neighbors of the vertex. Once again, this
demonstrates the importance of locality in paleoclimate Markov random
field structures. We note, however, that large improvements are still
recorded at locations with a larger number of neighbors. The larger
neighborhoods may represent real geophysical structures, in which case
the reconstruction may benefit from the flexibility of the model. These
edges may also be spurious. We note, however, that the presence of an
edge $(i,j)$ in the graph simply does not force the corresponding entry
$\omega_{ij}$ in the precision matrix to be zero. When $\Omega$ is
estimated in accordance with the graph, $\omega_{ij}$ can still be
very small. Large improvements are therefore possible when spurious
edges are present in the graph.

%s6 #&#
\section{Concluding remarks}
The main objective of the paper was to explore the efficacy of recent
advances in the theory of graphical models and high-dimensional
inference for statistical paleoclimate reconstructions. Markov random
fields provide a sparse representation of the precision matrix of
spatial fields, and thus achieve the dimension reduction that is often
necessary in high-dimensional settings.

We explored two families of methods to estimate the graphical structure
of climate fields: a neighborhood approach and a $\ell_1$ penalized
inverse covariance estimation approach. In neighborhood graphs, each
vertex is only connected to its \mbox{immediate} neighbors, reflecting the
fact that variables at two locations are expected to be independent
given the temperature in a geographical neighborhood. The size of the
neighborhoods can be chosen from the data by cross-validation. The
$\ell_1$ method, in contrast, provides more flexibility to represent
the spatial heterogeneities of geophysical fields (e.g., land/ocean
contrasts, topographical boundaries, teleconnection patterns), which
would in general be difficult with parametric (e.g., Mat\'{e}rn family)
covariance functions. The GraphEM algorithm was subsequently tested on
pseudoproxy data. We also proposed a block bootstrap method to
internally estimate the uncertainties in the reconstructions performed
using GraphEM and RegEM-TTLS.

Our experiments show that the GraphEM approach gives consistently
better reconstructions than the frequently used RegEM-TTLS [see, e.g.,
\citeauthor{Mann08a} (\citeyear{Mann08a,Mann09})] almost uniformly over space. We show that
Gaussian Markov random fields yield demonstrably improved estimates of
the underlying spatio-temporal process, which we tied to the sparsity
of the estimated covariance model. A caveat of the $\ell_1$ method is
the tendency to sometimes detect spurious edges in the graph, that is,
to detect relationships that arise from the presence of noise, instead
of physical links between the temperature field and the proxies (or
pseudoproxies) that derive from it. This is to be expected due to the
signal to noise relationship in the data, and is inherent in all
statistical and signal processing recovery techniques. Further
constraints on the graph can naturally be added to ensure that the
graphs selected by the \textit{graphical lasso} retain a high degree
of locality. In contrast, neighborhood graphs seem to provide an
adequate approximation to the conditional structure of the
temperature/proxy field, independently of the level of noise present in
the data. The size of the neighborhoods can also be chosen from the
data so as to minimize the prediction error. As we demonstrate in our
simulations, neighborhood graphs perform well and can be used in
situations where there is less hope of discovering the graphical
structure of the field from the data. We also observed that most
locations in the graphs estimated using the $\ell_1$ method are
connected to geographically close locations. Locality is therefore an
important feature in paleoclimate graphs. We also note that most
proxies have no proxy neighbors in graphs estimated from the data,
suggesting that proxies are independent of each other given the
temperature data.

Finally, and although we were primarily motivated by paleoclimate
applications and the use of the EM algorithm in this context, it is
worth pointing out that graphical models are also applicable within
Bayesian CFR methods [e.g., \citeauthor{TingleyHuybers10a} (\citeyear{TingleyHuybers10a,TingleyHuybers10b})] and beyond the confines of
climate science. GraphEM as described here provides a useful addition
to the RegEM framework, one that will be applicable to any
high-dimensional imputation problem, and one that can be used in tandem
with other $\ell_2$ regularization approaches, especially
data-adaptive ones. Future work will extend the use of Gaussian Markov
random fields as process models for geophysical fields, in tandem with
hierarchical models.

\begin{appendix}
\section{Description of the GraphEM algorithm}\vspace*{-6pt}\label{appalgo}
\begin{algorithm}[H]
\caption{The graphical EM algorithm (GraphEM)}\label{algoGraphEM}
\begin{center}\textbf{Input:} Incomplete $n \times p$ matrix $X$, graph $G$.\end{center}
\begin{algorithmic}[1]
\STATE Initialize $X^{(0)}$ by replacing the missing values in $X$ by
the sample mean of each variable over the instrumental period;
\STATE Compute initial estimates $\mu^{(0)}$ and $\Sigma^{(0)}$ of
$\mu$ and $\Sigma$ by computing the sample mean and sample covariance
of $X^{(0)}$;
\STATE Initialize $i \leftarrow0$;
\STATE Initialize $\Sigma_G^{(0)} = \Sigma^{(0)}$;
\REPEAT
\STATE Compute $X^{(i+1)}$ by performing a linear regression of the
missing values on the available ones for each row of $X$, using the
current estimate $\mu^{(i)}$ of $\mu$ and the current graphical
estimate $\Sigma_G^{(i)}$ of $\Sigma$ [see (\ref{eqnregression})];
\STATE Compute $\mu^{(i+1)}$ by computing the sample mean of $X^{(i+1)}$;
\STATE Compute $\Sigma^{(i+1)}$ as in (\ref{eqnEMupdate});
\STATE Compute the new graphical estimate $\Sigma_G^{(i+1)}$ by
solving (\ref{eqngraphicalMLE}) with $S = \Sigma^{(i+1)}$, that is,
%e7.1 #&#
%
\begin{equation}
\label{eqngraphicalMstep} \Sigma_G^{(i+1)} = \mathop{\argmax\limits_{\Sigma= \Omega^{-1} >
0}}_{\Omega_{ij} = 0, (i,j) \notin E} \log\det\Omega- \tr\bigl(\Sigma
^{(i+1)} \Omega\bigr);
\end{equation}
\STATE$i \leftarrow i+1$;
\UNTIL{convergence}
\end{algorithmic}
\begin{center}\textbf{Output:} Completed matrix $\hat{X}$, estimate $\hat{\mu}$
of $\mu$, estimate $\hat{\Sigma}$ of $\Sigma$.\end{center}
\end{algorithm}

%
%s8 #&#
\section{Derivation of the Graph-EM algorithm}\label{appEM}

We follow the notation in Little and Rubin [\citet
{LittleRubin2002}]. The complete data belongs to a regular exponential
family given by a Gaussian Markov random field with graph $G = (V,E)$
(as compared to a complete model in the classical EM algorithm). The
sufficient statistics are given by
%e8.1 #&#
%
\begin{equation}
\mathcal{S} = \Biggl(\sum_{i=1}^n
y_{ij}, j=1, \ldots, k ;\sum_{i=1}^n
y_{ij} y_{ik}\mbox{ with } (j,k) \in E \Biggr).
\end{equation}
Let $\theta^{(t)} = (\mu^{(t)}, \Sigma^{(t)})$ denote the current
estimate of the parameters. The E-step is given as follows:
%e8.2 #&#
%
\begin{equation}
\label{eqnE1} E \Biggl[\sum_{i=1}^n
y_{ij} \bigg| Y_{\mathrm{obs}}, \theta^{(t)} \Biggr] = \sum
_{i=1}^n y_{ij}^{(t)},
\qquad j=1, \ldots,k
\end{equation}
and
%e8.3 #&#
%
\begin{equation}
\label{eqnE2} E \Biggl[\sum_{i=1}^n
y_{ij}y_{ik} \bigg| Y_{\mathrm{obs}}, \theta^{(t)}
\Biggr] = \sum_{i=1}^n
\bigl(y_{ij}^{(t)} y_{ik}^{(t)} +
c_{jki}^{(t)}\bigr), \qquad(j,k) \in E
\end{equation}
with
%e8.4 #&#
%
\begin{equation}
y_{ij}^{(t)} = \cases{ y_{ij}, &\quad when
$y_{ij}$ is observed,
\vspace*{3pt}\cr
E \bigl[y_{ij} \mid
y_{\mathrm{obs}, i}, \theta^{(t)} \bigr], &\quad when $y_{ij}$ is
missing}
\end{equation}
and
%e8.5 #&#
%
\begin{equation}
\qquad c_{jki}^{(t)} = \cases{ 0,\qquad\mbox{if at least one of the $y_{ij}$ or $y_{ik}$ is observed,}
\vspace*{3pt}\cr
0,\qquad\mbox{if $j \perp_G k \mid\mathrm{obs}, i$},
\vspace*{3pt}\cr
\cov\bigl[y_{ij},
y_{ik} \mid y_{\mathrm{obs},i}, \theta^{(t)} \bigr],
\cr
\hspace*{33pt}\mbox{if both $y_{ij}$ and $y_{ik}$ are missing and $j \not
\perp_G k \mid\mathrm{obs}, i$,}}
\end{equation}
where $j \perp_G k$ means that $j$ and $k$ are separated in the graph
$G$ [see, e.g., \citet{lauritzen}, Example 3.2]. At a first
glance, it would appear as if there is little difference between the
treatment in the graphical vs. the complete case. A closer look reveals
that there are some notable differences, the\vspace*{1pt} first being in the
calculation of the sufficient statistics. Second, note that the
definition of $y_{ij}^{(t)}$ and $c_{jki}^{(t)}$ below are different:
$y_{ij}^{(t)}$ when $y_{ij}$ is missing is given as follows:
%e8.6 #&#
%
\begin{eqnarray}
&&  E \bigl[y_{ij} \mid y_{\mathrm{obs}, i}, \theta^{(t)} \bigr]
\nonumber\\[-8pt]\\[-8pt]\nonumber
&&\qquad =
\mu_j^{(t)} + \bigl(\Sigma_{j,\mathrm{obs}}^G
\bigr)^{(t)} \bigl[ \bigl(\Sigma_{\mathrm{obs},\mathrm{obs}}^G
\bigr)^{(t)} \bigr]^{-1} \bigl(y_{\mathrm{obs}, i} -
\mu_{\mathrm
{obs}}^{(t)} \bigr),
\end{eqnarray}
where $ (\Sigma^G )^{{(t)}}$ corresponds to a graphical
covariance matrix $\Sigma$. When both $y_{ij}$ and $y_{ik}$ are
missing and $(j,k) \in E$,
%e8.7 #&#
%
\begin{eqnarray}
&& \cov\bigl[y_{ij} y_{ik} \mid y_{\mathrm{obs}, i}, \theta
^{(t)} \bigr]
\nonumber\\[-8pt]\\[-8pt]\nonumber
&&\qquad = \bigl(\Sigma_{jk}^G
\bigr)^{(t)} - \bigl(\Sigma_{\{j,k\}, \mathrm{obs}}^G
\bigr)^{(t)} \bigl[ \bigl(\Sigma_{\mathrm{obs},\mathrm{obs}}^G
\bigr)^{(t)} \bigr]^{-1} \bigl(\Sigma_{\mathrm{obs}, \{j,k\}}^G
\bigr)^{(t)}.
\end{eqnarray}
Note, however, that $\cov[y_{ij} y_{ik} \mid y_{\mathrm
{obs}, i}, \theta^{(t)} ] = \Sigma_{jk \mid\mathrm
{obs},i}^{(t)}$. Thus,
%e8.8 #&#
%
\begin{equation}
c_{jki}^{(t)} = \Sigma_{jk \mid\mathrm{obs},i}^{(t)} = 0\qquad\mbox{if } j \perp_G k \mid\mathrm{obs},i.
\end{equation}
The M-step in the GraphEM algorithm therefore consists of using the
sufficient statistics for the complete data derived in (\ref{eqnE1})
and (\ref{eqnE2}) to determine the graphical mle. In particular, the
estimate of the mean parameter is given by the sample mean and the
estimate of the graphical covariance is given in equation (\ref
{eqngraphicalMstep}).

%s9 #&#
\section{Nonparametric bootstrap}\label{appbootstrap}
\begin{algorithm}[H]
\caption{RegEM/GraphEM Uncertainty quantification (nonparametric bootstrap)}\label{algobootstrap}
\begin{center}\textbf{Input:} Incomplete $n \times p$ matrix $X$ containing $n_i$
years of instrumental data, number of bootstrap samples $N > 1$,
blocksize $b$.\end{center}
\begin{algorithmic}[1]
\FOR{$i=1, \ldots, N$}
\STATE Construct a bootstrap sample $X_{\mathrm{boot},i}$ by sampling
with replacement $\lceil n_i/b\rceil$ blocks of size $b$ from the
lines of $X$ in the instrumental period, and $\lceil(n-n_i)/b\rceil$
blocks of size $b$ from lines in the rest of the matrix;
\STATE Reconstruct the missing values in $X_{\mathrm{boot},i}$ using
RegEM/GraphEM. The algorithm outputs estimates $\mu_{\mathrm{boot},
i}$, $\Sigma_{\mathrm{boot},i}$ of the mean and covariance matrix of
the field;
\STATE Obtain $\hat{X}_i$ by reconstructing the missing values in $X$
by performing the regression step of RegEM/GraphEM starting with $\mu
_{\mathrm{boot}, i}$ and $\Sigma_{\mathrm{boot}, i}$;
\STATE For each line $(x_m, x_a)$ of $\hat{X}_i$ that originally
contained missing values, add a noise realization to $x_m$ from the
conditional distribution of $x_m\mid x_a$, where we assume $(x_m,x_a)
\sim
N(\mu_{\mathrm{boot}, i}, \Sigma_{\mathrm{boot},i})$.
\ENDFOR
\end{algorithmic}
\begin{center}\textbf{Output:} Ensemble of $N$ reconstructions $\hat{X}_1, \ldots,
\hat{X}_N$ of the incomplete field $X$.\end{center}
\end{algorithm}
\end{appendix}

\section*{Acknowledgement}
We wish to thank Martin Tingley for useful
comments and suggestions that have greatly improved the paper.

%suskaldyti doi

% imsref loaded by linak, 2014-11-26 12:50:32
%

\printaddresses
\end{document}